*Comment*

# Comments on "Mathematical Modeling of Current Source Matrix Converter with Venturini and SVM"

**Irfan Ahmad Khan[1,*] and Anshul Agarwal[2]**

[1,2] Electrical and Electronics Engineering, National Institute of Technology Delhi, Delhi-110040 (India); irfan00706@gmail.com

* Correspondence: irfankhan@nitdelhi.ac.in (I.K.)

**Abstract:** In this paper, authors want to comment on a recently published article describing the Mathematical Modeling of Current Source Matrix Converter (CSMC) with two modulation strategies, namely: Venturini and Space Vector Modulation (SVM). Reported flaws are broadly classified into two (2) categories; namely: **(1) Technical Flaws** and **(2) Rubric and Grammar Flaws.** This paper not only reports these flaws but also provides suggestive rectifications for correction and improvement of the published article.

## 1. Introduction

Frequency converters (FC) are the power electronic (PE) converters, which convert alternating current (AC) electrical power with specified parameters into AC electrical power with desired parameters [1-3]. The direct matrix converter (MC) is a static AC-AC converter used for converting AC voltage at fixed frequency directly to an adjustable voltage with adjustable frequency (AVAF), without requiring an intermediate DC stage [4-6].

## 2. Flaws and Suggestive Rectifications

In this paper an attempt has been made not only to notify the flaws but also to provide suggestive rectifications, which may lead to corrective measures for improving the quality of published article. These flaws are broadly classified into two (2) categories; namely: (1) Technical Flaws and (2) Rubric and Grammar Flaws. Following subsections are presenting the flaws along with suggestive rectification measures.

2.1. Technical Flaws

There are following four (4) technical flaws reported in the published article:

2.1.1. Maximum Voltage Gain for the MC with SVM

In conclusion section of the published article, the maximum voltage gain is mentioned as **8.66** for SVM [1]. The authors want to report that for a 3x3 MC, it should be **less than unity (< 1)**.

The suggestive correction is that the maximum voltage gain = $\sqrt{3}/2 \approx 0.866$   [5,7]

2.1.2. Range of Voltage Gain for the MC with Venturini Modulation Strategy

In section 3 of the published article, the range of voltage gain is mentioned as:

q = $U_{Om}$ / $U_{Im}$ is the voltage gain: **0 < q < 0** [1]; i.e. zero to **zero**.

The authors want to report that, it should be greater than zero and less than **half (< 0.5)**.

The suggestive correction is that the range of voltage gain should be 0 < q < **0.5**.

2.1.3. Common-Emitter (CE) Topology for bi-directional IGBT switch used in MC

In section 2 of the published article, main topologies for bi-directional switch used in MC are presented in Figure 2. The common-emitter (CE) configuration of the IGBT switches is shown in Figure 2a[1].

The authors want to report that, the presented Figure 2a[1] **is wrong**. It should be for the common-collector (CC) configuration of the IGBT switches.

The suggestive correction for the CE configuration of the IGBT switches is shown in following Figure 1.

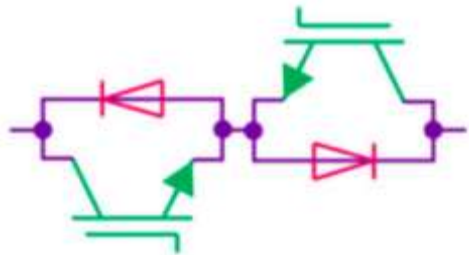

**Figure 1.** The common-emitter (CE) configuration of the IGBT switches [8].

2.1.4. Common-Collector (CC) Topology for bi-directional IGBT switch used in MC

In section 2 of the published article, main topologies for bi-directional switch used in MC are presented in Figure 2. The common-collector (CC) configuration of the IGBT switches is shown in Figure 2b[1].

The authors want to report that, the presented Figure 2b[1] **is wrong**. It should be for the common-emitter (CE) configuration of the IGBT switches.

The suggestive correction for the CC configuration of the IGBT switches is shown in following Figure 2.

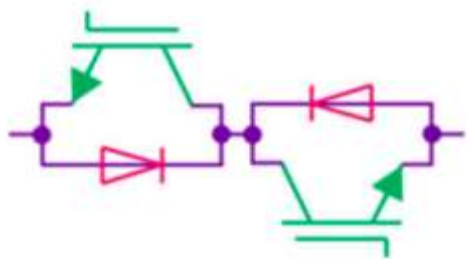

**Figure 2.** The common-collector (CC) configuration of the IGBT switches [8].

2.2. Rubric and Grammar Flaws

There are several Rubric and Grammar related flaws in the published article. Few of these flaws are mentioned as follows:

- In section 2 of the published article, first line states that "The schematic diagram of the current **PM** is shown in Figure 1b"[1]. The abbreviation **"PM"** is not defined anywhere in the published article.
- In last para of page 2 of the published article, second line states that "to model CSMC with **tow** control strategy….."[1]. The word **"tow"** should be replaced by **two** (2) in the published article.

**5. Conclusions**

In this comment paper, authors reported major flaws in a recently published article describing the Mathematical Modeling of Current Source Matrix Converter (CSMC) with two modulation strategies, namely: Venturini and Space Vector Modulation (SVM). Reported flaws have been presented for two (2) broad categories; namely: **(1) Technical Flaws** and **(2) Rubric and Grammar Flaws.** This comment paper not only reported these flaws but also provided suggestive rectifications for correction and improvement of the published article.

**Funding:** This research received no external funding.

**Conflicts of Interest:** The authors declare no conflict of interest.

**References**

1. Chodunaj, M.; Szcze´sniak, P.; Kaniewski, J. Mathematical Modeling of Current Source Matrix Converter with Venturini and SVM. *Electronics* **2020,** 9, 558.
2. Agarwal, A.; Agarwal, V.; Khan, I.A. FPGA Based Sinusoidal Pulse Width Modulated Frequency Converter. Proceedings of the IEEE 1st International Conference on Power Electronics, Intelligent Control and Energy Systems (ICPEICES 2016), Delhi, India, July 2016.
3. Khan, I.A.; Agarwal, A. Delta and Adaptive Delta Modulated Single Phase AC/AC Converter. Proceedings of the IEEE 1st International Conference on Power Electronics, Intelligent Control and Energy Systems (ICPEICES 2016), Delhi, India, July 2016.
4. Agarwal, A.; Khan, I.A.; Agarwal, V. FPGA Based Direct Matrix Converter: The Harmonic Analysis with Three Modulation Techniques. Proceedings of the IEEE Region 10 International Conference on the theme - Technologies for Smart Nation (TENCON 2016), M.B.S., Singapore, Nov. 2016.
5. Ali, M.; Iqbal, A.; Khan, M.R.; Ayyub, M.; Anees, M.A. Generalized Theory and Analysis of Scalar Modulation Techniques for a m × n Matrix Converter. *IEEE Trans. Power Electron.* **2017**, 32, 4864–4877.
6. Khan, I.A.; Agarwal, A. FPGA Based Adaptive Delta Modulated Direct Matrix Converter. Proceedings of the IEEE 7th Power India International Conference (PIICON 2016), Bikaner, India, Nov. 2016.
7. Szcze´sniak, P.; Kaniewski, J. Hybrid Transformer with Matrix Converter. *IEEE Trans. Power Electron.* **2016**, 31, 1388–1396.
8. Szcze´sniak, P. Challenges and Design Requirements for Industrial Applications of AC/AC Power Converters without DC-Link. *Energies* **2019**, 12, 1581.